\documentclass[aps,prl,twocolumn,superscriptaddress,showkeys,showpacs]{revtex4}

\usepackage{graphicx}
\usepackage{dcolumn}
\usepackage{bm}

\begin{document}


\title{Magnetic double gradient instability and flapping waves in a current sheet}


\author{N. V. Erkaev}
 \affiliation{Institute of Computational Modelling, Russian Academy of Sciences,
 } \affiliation{Siberian Federal
University, Krasnoyarsk, Russia}

\author{V. S. Semenov}

\affiliation{Institute of Physics, State University of St.
 Petersburg, St. Petersburg, Russia}

 \author{H. K. Biernat}

 \affiliation{Space Research Institute, Austrian Academy of
 Sciences, Graz, Austria}
\affiliation{Institute of Physics, University of Graz,
 Graz, Austria}


\date{\today}

\begin{abstract}
A new kind of magnetohydrodynamic instability and waves are analyzed
for a current sheet in the presence of a small normal magnetic field
component varying along the sheet. These waves and instability are
related to existence of two gradients of the tangential ($B_\tau$)
and normal ($B_n$) magnetic field components  along the normal
($\nabla_n B_\tau $) and tangential ($\nabla_\tau B_n $)  directions
with respect to the current sheet. The current sheet can be stable
or unstable if the multiplication of two magnetic gradients  is
positive or negative. {In the stable region, the ``kink''-like wave
mode is interpreted as so called flapping waves observed in the
Earth's magnetotail current sheet}. The ``kink'' wave group velocity
estimated for the Earth's current sheet is of the order of a few
tens kilometers per second. This is in good agreement with the
observations of the flapping motions of the magnetotail current
sheet.
\end{abstract}


\maketitle

\section{Introduction}

Thin current layers are typical structures in the Heliosphere,
including the solar corona, solar wind and planetary magnetospheres.
We address some of the magnetohydrodynamic aspects concerning the
stability of current layers which are still poorly understood. In
particular, CLUSTER observations in the Earth's magnetotail current
sheet indicated the appearance of strong wave perturbations
propagating across the current sheet. Many event studies indicated
very large current sheet variations and a predominant wave
propagation in the transverse direction with respect to the magnetic
field plane. The existence of such kind of waves associated with
flapping motions was confirmed in many statistical studies
\cite{Sergeev03,Sergeev04,Sergeev06,Runov05_a,Runov05_b,Runov06,Petruk}
which allowed one to identify them as the ``kink''-like
perturbations. The plasma sheet flapping observations are
interpreted as crossings of a quasi-periodic dynamical structure
produced by almost vertical slippage motion of the neighboring
magnetic flux tubes. The frequency of the flapping motions,
estimated from observations is $w_f \sim$
 0.035 s$^{-1}$ \citep{Sergeev03}.
For a majority of the observed events \citep{Runov05_a}, a group
speed of the flapping waves was found to be in the range of a few
tens (30--70) kilometers per second. The wavelengths and spatial
amplitudes are estimated to be of the order of 2 -– 5 $R_E$ ($R_E$
is the Earth's radius) \citep{Petruk}.

A preferential appearance of one (``kink''-like) mode of the
flapping motion was reported by \citep{Sergeev06}. CLUSTER
observations give rise to the assumption that the flapping motions
are notably more frequent in the central part of the tail than near
the flanks. In the near-flank tail regions the motions of flapping
waves are predominantly from the center to the flanks
\citep{Sergeev04}. These experimental results confirm an internal
origin of the flapping motions, due to some processes (like magnetic
reconnection) localized deep inside the magnetotail. On the basis of
CLUSTER observations of reconnection events, a relationship between
the flapping motion and the reconnection process was investigated by
\citep{Lait}. During the reconnection events the current sheet
exhibits strong flapping motions that propagate towards the flank of
the tail.

With regard to a theoretical aspect of the problem,  the
Ballooning-type mode in the curved current sheet magnetic field was
claimed to be able to propagate azimuthally in flankward directions
from the source \citep{Golov}. This ballooning theory was applied in
the WKB approximation implying the condition that the wave length
scale is much less than the curvature radius. This condition can
hardly be fulfilled in the plasma sheet with a small normal
component of the magnetic field. Another point is that according to
the theory of \citep{Golov}, both ``kink''-like and ``sausage''-like
deformations of the current sheet are equally possible, and the
question arises about a reason, why the observed flapping
perturbations of the current sheet are mainly associated with the
``kink''-like wave modes.

In this paper, we propose a new approach to explain the existence of
the ``kink''-like flapping wave oscillations propagating across the
current sheet. In a framework of a rather simple magnetohydrodynamic
consideration, we elucidate a physical reason of the flapping wave
oscillations of the current sheet, which is related with gradients
of the tangential and normal magnetic field components with respect
to the normal and tangential directions, respectively.
\section{Statement of problem}

A geometrical situation of the problem and coordinate system are
illustrated in Fig.~1. We apply a system of incompressible ideal
magnetohydrodynamics for nonstationary variations of plasma sheet
parameters
\begin{eqnarray}
\rho \left( \frac{\partial {\bf V}}{\partial t} + {\bf V}\cdot
\nabla {\bf V} \right) + \nabla P = \frac{1}{4\pi} {\bf B}\cdot
\nabla{\bf B} , \label{MHD1}\\
\frac{\partial {\bf B}}{\partial t} + {\bf V}\cdot \nabla  {\bf B} =
{\bf B}\cdot\nabla{\bf V},\label{MHD2}\\
\nabla\cdot{\bf V} =0, \quad \nabla\cdot {\bf B} =0.\label{MHD3}
\end{eqnarray}
Here ${\bf V}, {\bf B}, \rho, P$ are the velocity, magnetic field,
density and total pressure, respectively. The total pressure is
defined as the sum of the magnetic and plasma pressures.
 { We consider
specific wave perturbations propagating across the magnetic field
lines, which are much slower than the magnetosonic modes. In this
case the incompressible approximation seems to be appropriate. }

{ We focus our study on the very slow wave modes existing only in
the presence of a gradient of the $B_z$ component in the magnetotail
current sheet along the $x$ direction.}
 The background conditions are considered to be rather simple with
 a slow dependence of the $B_z$ component on the $x$ coordinate
\begin{eqnarray}
B_x = B^* b_x(\bar{z}), \quad B_z = \varepsilon B^* b_z(\bar{x}),
\quad B_y =0, \nonumber \\ {\bf V} =0, \quad \bar{y} = y/\Delta,
\quad \bar{z} = z/\Delta, \quad \bar{x} = x/L_x.
\end{eqnarray}
Here $\Delta$ is a thickness of the current sheet, and $L_x$ is a
length scale of the $B_z$ variation along the current sheet.

 We introduce normalized small perturbations marked by sign
``tilde'' which are considered to be functions of time and two
spatial coordinates ($y, z$)
\begin{eqnarray}
B_x = B^*(\tilde b_x + b_x(\bar{z})), \quad B_y =
\varepsilon B^* \tilde b_y , \nonumber \\
B_z = \varepsilon B^*(b_z(\bar{x}) + \tilde b_z), \quad P = P_0+\tilde P B^{*2}/(4\pi),\nonumber \\
\quad V_x = \tilde v_x V_A, \quad V_y = \tilde v_y V_A, \quad V_z =
\tilde
v_z V_A, \nonumber\\
\bar{t} = t V_A/\Delta, \quad V_A = B^*/\sqrt{4\pi\rho^*}, \quad \nu
=\Delta/L_x.
\end{eqnarray}
Here $P_0$ is the background total pressure, the parameter
$\varepsilon$ means the ratio of the background normal and maximal
tangential components of the magnetic field, and the parameter $\nu$
{ characterizes the gradient of the normal magnetic field component.
For the background conditions considered in our model
($B_z(\bar{x}), B_x(\bar{z})) $, equation $\nabla\cdot{\bf B} = 0$
is fulfilled for arbitrary independent parameters $\varepsilon$ and
$\nu$.}

{ Linearizing Eqs.~(\ref{MHD1}--\ref{MHD3}) for the normalized
perturbations, neglecting high order terms $\sim \nu^2 \varepsilon$,
and $\sim \varepsilon^2$, we assume $\nu \gg \varepsilon$ and retain
the main term $\sim \nu\varepsilon$. }

Substituting Fourier harmonics ($\propto exp(i \bar{\omega}\bar{t} -
i \bar{k} \bar{y}) )$, we obtain finally a system of equations for
Fourier amplitudes
\begin{eqnarray}
i\bar{\omega} \tilde v_x = \varepsilon \left(\tilde b_z \frac{d
b_x}{d \bar{z}} + b_z \frac{d \tilde b_x}{d \bar{z}} \right),\label{Vx}\\
i \bar{\omega}  \tilde v_y - i \bar{k} \tilde P =0, \quad i
\bar{\omega} \tilde v_z + \frac{d \tilde P}{d \bar{z}} = \varepsilon
\nu
\tilde b_x \frac{ d b_z}{d \bar{x}},\label{Vz}\\
i \bar{\omega} \tilde b_z - b_z \frac{d \tilde v_z}{d \bar{z}}
+\nu\tilde v_x \frac{d b_z}{d \bar{x}} = 0, \quad i\bar{\omega}
\tilde b_y -  b_z \frac{d \tilde v_y}{d \bar{z}}=0,\label{byz}\\
 i\bar{\omega} \tilde b_x + \frac{d b_x}{d\bar{z}} \tilde v_z=0,
\quad -i \bar{k} \tilde v_y + \frac{d \tilde v_z}{d\bar{z}}=0.
\label{bxvy}
\end{eqnarray}
{ In this system of equations the derivative $d b_z/d\bar{x}$ is
assumed to be constant, and all other quantities are considered to
be not dependent on the $x$ coordinate. Therefore Eqs.~(\ref{Vx}
-\ref{bxvy}) are treated as a system of ordinary differential
equations with respect to the $\bar{z}$ coordinate}. Excluding
$\tilde b_x$ and $\tilde b_z$ in Eq.~(\ref{Vx}), we derive
\begin{eqnarray}
\tilde v_x \left(-\bar{\omega}^2+ U(\bar{z})\right)=0, \quad
 U(\bar{z}) = \varepsilon\nu\frac{d b_x}{d \bar{z}}\frac{d  b_z}{d \bar{x}}.\label{Vx1}
\end{eqnarray}
Generally, for a nonconstant $U(\bar{z})$, Eq.~(\ref{Vx1}) yields
$\tilde v_x =0.$

From  Eqs.~(\ref{Vz}--\ref{bxvy}),  we finally obtain a second order
ordinary differential equation for the $\tilde v_z$ velocity
perturbation
\begin{eqnarray}
\frac{d^2 \tilde v_z}{d \bar{z}^2} +\bar{k}^2 \tilde v_z
\left(\frac{U(\bar{z})}{\bar{\omega}^2} -1\right) = 0.
\label{spectral}
\end{eqnarray}

Further for simplicity we consider a piecewise constant function
$U(\bar{z})$
\begin{eqnarray}
U(\bar{z}) = \varepsilon \nu , \, \, -1 \leq \bar{z} \leq 1; \quad
U(\bar{z}) =  0, \quad |\bar{z}| > 1, \label{U(z)}
\end{eqnarray}
which means that the current density is assumed to be constant
within the current sheet.

\section{Results}

A choice of the piecewise constant function $U(\bar{z})$ allows us
to find analytical solutions which are of two kinds, ``kink''-like
and ``sausage''-like modes. The ``kink''-like mode is characterized
by displacement of the current sheet center, and even function
$\tilde v_z(\bar{z})$
\begin{eqnarray}
\tilde v_z = C \exp(-\bar{k} (|\bar{z}|-1)), \quad |\bar{z}| > 1;\\
\tilde v_z = D \cos(\lambda \bar{z}), \quad \lambda = \bar{k}
\sqrt{\varepsilon \nu/\bar{\omega}^2 -1}, \quad |\bar{z}| \leq 1 .
\label{kink_sol}
\end{eqnarray}
An odd function $\tilde v_z(\bar{z})$ is relevant to the
``sausage''-like mode characterized by variations of the thickness
of the current layer without a displacement of its center
\begin{eqnarray}
\tilde v_z = C \exp(-\bar{k} (\bar{z}-1)), \quad \bar{z} > 1;
\nonumber \\
\tilde v_z = -C \exp(\bar{k} (\bar{z}+1)), \quad \bar{z} < -
1; \nonumber \\
\tilde v_z = D \sin(\lambda \bar{z}), \quad \lambda = \bar{k}
\sqrt{\varepsilon \nu/\bar{\omega}^2 -1}, \quad |\bar{z}| \leq 1 .
\label{sausage_sol}
\end{eqnarray}
Applying continuity conditions for $\tilde v_z$ and the first
derivative $d\tilde v_z / d\bar{z}$ at the current layer boundaries,
we obtain algebraic system corresponding to the ``kink'' mode
\begin{eqnarray}
C = D \cos(\lambda), \quad \bar{k}C = \lambda D\sin(\lambda ),
\label{kink_eqs}
\end{eqnarray}
and also we find a system for the ``sausage'' mode
\begin{eqnarray}
C = D \sin(\lambda), \quad -\bar{k}C = \lambda D \cos(\lambda ).
\label{sausage_eqs}
\end{eqnarray}
Setting the determinants to vanish, we derive two equations
corresponding to the ``kink'' and ``sausage'' modes, respectively
\begin{eqnarray}
\tan(\lambda) = \frac{\bar{k}}{\lambda } \,(\mbox{``kink''}); \,
\tan(\lambda) = -\frac{\lambda}{\bar{k}}
\,(\mbox{``sausage''}).\label{lambda_eqn}
\end{eqnarray}
These equations have  discrete sequences of roots $\lambda_1,
\lambda_2, ...\lambda_n,...$ . The main root is the minimal
$\lambda$ which corresponds to the maximal frequency.

 By numerical solving these equations, we obtain two main roots
$\lambda_{k,s}$ which determine the dimensional frequencies
$\omega_{k,s}$ as functions of wave number for the ``kink'' and
``sausage'' modes
\begin{eqnarray}
\omega_{k,s} =  \omega_f \frac{k\Delta}{\sqrt{k^2\Delta^2 +
\lambda_{k,s}^2}}, \quad \omega_f =
\sqrt{\frac{1}{4\pi\rho}\frac{\partial B_x}{\partial
z}\frac{\partial B_z}{\partial x}}. \label{omega_ks}
\end{eqnarray}
Here $\omega_f$ means a characteristic flapping frequency
proportional to the square root of the multiplication of two
gradients of the background magnetic field components, $\partial
B_x/\partial z$ and $\partial B_z /\partial x$. The dimensionless
functions $\omega_{k,s}/\omega_f$ are presented at the top panel in
Fig.~2. Frequencies are monotonic functions of wave number, and they
increase to the maximal asymptotic value $\omega_f$ for $k\Delta
\rightarrow \infty$. The group wave velocity
 is shown in Fig.~2 as functions of wave
number (the second panel). It decreases monotonically to zero for
increasing wave numbers.

The flapping wave perturbations become unstable when the
multiplication of two magnetic gradients becomes negative.
In particular, for the Earth's plasma sheet this condition
corresponds to the case of decreasing $B_z$ component towards Earth.
The growth times of the instability for the ``kink'' and ``sausage''
modes are given by formulas
\begin{eqnarray}
\tau_{k,s} = \tau_f\frac{\sqrt{\lambda_{k,s}^2+k^2
\Delta^2}}{k\Delta} , \,\,\, \tau_f =1 /
\sqrt{\frac{-1}{4\pi\rho}\frac{\partial B_x}{\partial
z}\frac{\partial B_z}{\partial x}}.
\end{eqnarray}
The instability growth times ($\tau_{k,s}/\tau_f$) are shown in
Fig.~2 (bottom panel) as functions of wave number for the two wave
modes. One can see from the figure that the unstable ``kink'' mode
develops much faster than the sausage mode. In particular, for
$k\Delta = 0.7$ the ratio of growth times is $\tau_{s}/\tau_{k} =
2$. Fig.~3 illustrates a perturbation of the current sheet and
 the directions of plasma motion corresponding to the ``kink''
 mode flapping.

 A qualitative explanation of the flapping instability and waves
 corresponding to the obtained solution is
 the following.
Let us consider a  plasma element of a unit volume at the center of
the current layer as shown in Fig.~4.
 Along the
$z$ direction the resulting force $F_z$ acting on this plasma
element is a difference of two forces caused by the magnetic stress
and the total pressure gradient. In equilibrium state, the resulting
force $F_z$ vanishes, and the total pressure gradient compensates
the magnetic stress
\begin{eqnarray}
\frac{\partial P}{\partial z} = \frac{1}{4\pi}B_x \frac{\partial
B_z}{\partial x}.
\end{eqnarray}
In the new position of the magnetic tube element, the resulting
force will be
\begin{eqnarray}
F_z = -\frac{1}{4\pi} B_x(\delta z)\frac{\partial B_z}{\partial x} =
 -\frac{1}{4\pi} \delta z \left(\frac{\partial B_x}{\partial
z}\frac{\partial B_z}{\partial x}\right)_{z=0}.
\end{eqnarray}
This force accelerates plasma in the $z$ direction
\begin{eqnarray}
\rho \frac{\partial^2 \delta z}{\partial t^2} = - \delta
z\frac{1}{4\pi} \frac{\partial B_x}{\partial z}\frac{\partial B_z}
{\partial x}.
\end{eqnarray}
This equation yields the characteristic flapping frequency
$\omega_f$ which is proportional to the square root of the gradients
of the magnetic field components. This qualitative explanation of
the instability is illustrated in Fig.~4 where panels (a) and (b)
correspond to the stable and unstable situations, respectively.

For example, we estimate this frequency for the parameters which
seem to be reasonable for the conditions of the current sheet in the
Earth's magnetotail,
\begin{eqnarray}
B_x = 20 \,\mbox{nT}, \, B_z = 2 \,\mbox{nT}, \, \Delta \sim R_E, \,
n_p = 0.1 \,\mbox{cm}^{-3}, \nonumber\\ k\Delta =0.7, \,
\partial B_z/\partial x \sim B_z / L_x, \, L_x \sim 5 R_E.
\label{par_sheet}
\end{eqnarray}
For these parameters we find the characteristic flapping frequency
$\omega_f \sim$ 0.03 s$^{-1}$, and also the group velocity $V_g =
60\, $km/s.

\section{Summary}

The flapping instability and waves are analyzed for a current sheet
in a presence of two gradients of the $B_x$ and $B_z$ magnetic field
components along the $z$ and $x$ directions, respectively. These
both gradients play a crucial role for the stability of the current
sheet. The instability occurs in the regions of the current layer
where the multiplication of two gradients is negative. { In
particular, the instability can arise in a vicinity of a localized
thinning of the current sheet (Fig.~4b)}.
In  stable regions, the flapping waves are associated with
 the { so called ``Bursty Bulk Flows''
or BBF's  \citep{Sergeev06}, which are the magnetic tubes rapidly
moving through the center of the current sheet towards the Earth.
These BBF's are considered to be the sources of the flapping wave
oscillations propagating from the center  of the current sheet
towards the flanks in the $\pm y$ directions.}

The analytical solution is obtained for the simplified model of the
current layer with a constant current density.  The frequency and
the growth rate for the ``kink'' mode are found to be much larger
than those for the ``sausage'' mode. For both modes, the frequencies
are  monotonic increasing functions of the wave number. The
corresponding wave group velocities are decreasing functions of the
wave number, and they vanishes asymptotically for high wave numbers.

For the typical parameters of the Earth's current sheet, the group
velocity of the ``kink''-like mode is estimated as a few tens of
kilometers per second that is in good agreement with the CLUSTER
observations. A strong decrease of the group velocity for high wave
numbers means that the small scale oscillations propagate much
slower than the large scale oscillations. Because of that, the
propagating flapping pulse is expected to have a smooth gradual
front side part, and a small scale oscillating backside part.

{ The neglected second order terms $O(\varepsilon^2)$ are
responsible for the small effects related to the Alfv\'en waves
propagating in the $z$ direction. These second order effects are
subjects for future study. For the double gradient flapping waves
studied in our model, magnetic tension is not pronounced, because
the flapping waves propagate in the direction perpendicular to the
plane of the background magnetic field lines. The magnetic field
planes are just shifting with respect to each other}.

\begin{figure}
\noindent\includegraphics[width=14pc]{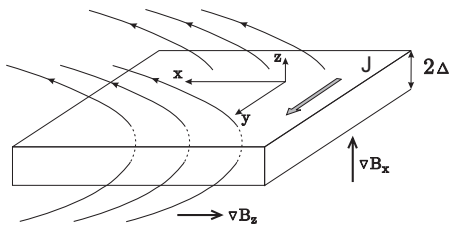}
 \caption{Geometrical situation of the problem}\label{Fig1}
\end{figure}
\begin{figure}
\noindent\includegraphics[width=16pc]{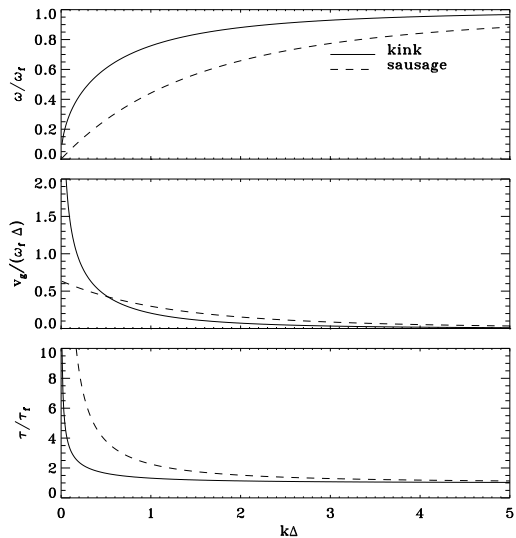}
 \caption{Frequency, group velocity, and instability growth time as functions
 of wave number for two wave modes}\label{Fig2}
\end{figure}
\begin{figure}
\noindent\includegraphics[width=15pc]{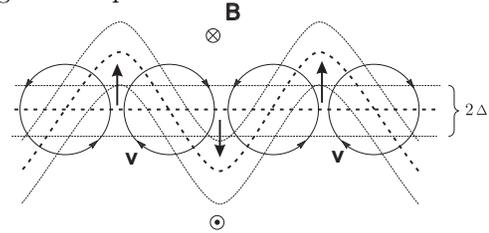}
 \caption{Illustration to the ``kink'' mode. Perturbation of the current sheet and
 the corresponding directions of plasma motion}\label{Fig3}
\end{figure}
\begin{figure}
\noindent\includegraphics[width=15pc]{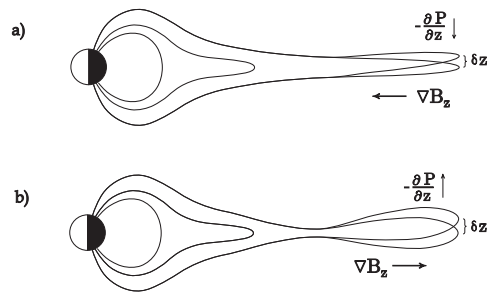}
\caption{Illustration to the ``kink'' flapping waves (a) and
instability (b) in cases of positive and negative gradient of $B_z$.
Displacements of the magnetic tubes are shown. }\label{Fig4}
\end{figure}

\begin{acknowledgments}
{\bf Acknowledgments}. {We thank Prof. V. Sergeev and
Dr.~I.~Kubyshkin for fruitful discussions and help in preparation of
the manuscript. This work is supported by RFBR grants N
07-05-00776-a, N 07-05-00135, by Programs 2.16 and 16.3 of RAS, and
by project P17100--N08 from the Austrian ``Fonds zur F\"orderung der
wissenschaftlichen Forschung'', and also by project I.2/04 from ``
\"{O}sterreichischer Austauschdienst''.}
\end{acknowledgments}

\end{document}